# Cooperative Adaptive Cruise Control with Variable Time Headway for Graceful Degradation under Fluctuating Network Quality of Service

Johannes Böhm[1], Eric Schöneberg[1], Kevin Schmidt[2], Wolfgang Fischer[2] and Daniel Görges[1]

***Abstract*—This paper proposes a dynamic distance adaptation for Cooperative Adaptive Cruise Control (CACC) under time-varying network conditions. When the Quality of Service (QoS) drops below a level required to maintain desired inter-vehicle distances, an online adaptation of the reference distances, reflected by a change of the time headway factor, becomes necessary. We present a control design algorithm realizing a graceful degradation, for which a distance control to a virtual preceding vehicle is introduced. Furthermore, the Integral Quadratic Constraints (IQC) framework is applied to guarantee robust stability of the time-varying system. The concept is validated in simulation and experimentally using small-scale test vehicles.**

## I. Introduction

Adaptive Cruise Control (ACC) is a well-established commercially available Advanced Driver Assistance System (ADAS). Besides its advantages of increasing the driver's comfort, it can also improve safety, traffic throughput and fuel consumption. Enabling communication between vehicles via Vehicle-to-Vehicle (V2V) technology and cloud-based control promises further enhancements in these aspects by allowing shorter inter-vehicle distances, leading to the concept of Cooperative Adaptive Cruise Control (CACC). However, networked control can be unreliable and scenarios involving communication degradation and time-varying Quality of Service (QoS), like communication delays, packet losses or complete connection losses, must be taken into account. A fixed-distance CACC guaranteeing stability and collision avoidance under all network conditions would be overly conservative. A more promising approach is to gracefully degrade the controller's performance online via adaptive inter-vehicle distances. This maintains stability and collision avoidance with only slight throughput and comfort reductions during low QoS while still outperforming ACC.

This work was supported by the Federal Ministry for Economic Affairs and Energy under grant no. 13IPC021.

[1]Böhm, Schöneberg, Görges are with the Department of Electrical and Computer Engineering, RPTU University Kaiserslautern-Landau, 67663 Kaiserslautern, Germany, {johannes.boehm, eric.schoeneberg, daniel.goerges}@rptu.de.

[2]Schmidt, Fischer are with the Corporate Research of Robert Bosch GmbH, 71272 Renningen, Germany, {kevin.schmidt4, wolfgang.fischer2}@de.bosch.com.

### A. Literature Review

Relevant research activities concerning ACC and CACC are the findings in, among others, [1], [2], [3] and [4]. In these works, it is pointed out that the commonly used condition of string stability for collision avoidance is not a sufficient condition. Other conditions, which include an externally positive behavior of the inter-vehicle's spacings, and systematic design procedures for CACC are presented which fulfill collision avoidance for constant vehicle-spacings with time-invariant time headway factors. These investigations will serve as basis for our research with adaptive, network-dependent time headway factors.

Considering network uncertainties, there are many CACC methods that mostly focus on time-invariant systems. A controller robust against uncertain communication delays was proposed in [5] using $\mu$ synthesis. The behavior for different constant time gaps was investigated, but no online adaptation was made. For complete connection losses with adaptive time headways, a controller which switches online between ACC and CACC when the communication is established or gets lost was designed in [6]. Also, a strategy introducing a "degraded CACC" as transitional mode was proposed [7]. However, all preceding concepts are limited to one-vehicle look-ahead situations, not exploiting the full potential of CACC. A lower-bound time headway for multiple-vehicle look-ahead architectures was derived in [8], but only for constant, uniform communication delays over the entire platoon. The authors of [9] proposed a controller that considers the condition of external positivity and adapts to uncertain vehicle parameters. Because it uses the same time headway for ACC and CACC, the inter-vehicle distance in CACC may be larger than required. In [10], an algorithm to adapt the time headway to the network conditions, especially packet losses, was presented and applied to a linear cooperative controller using the directly-preceding vehicle's data. Simulation confirmed an increased traffic flow and string stability for different simulation scenarios, but the stability of the adaptive controller was not proven formally.

### B. Contribution and Paper Overview

Based on the existing theory on ACC and CACC in [1], [2], [3] and [4] for time-invariant time headway factors, we investigate effects of time-varying network QoS and propose modifications to enable graceful degradation of the control algorithm's performance. Extending the velocity control in the CACC design, we present the control of the distance to a vehicle's virtual predecessor rather than the direct predecessor. This distance control becomes important when the network conditions degrade below a level requiring the distances between the vehicles to be enlarged, which will be reflected by a change of the vehicle's time headway factor. As internal stability of each local controlled vehicle is a necessary requirement, we demonstrate how to prove robust stability for the resulting linear parameter-varying (LPV) system with communication delay using the framework of Integral Quadratic Constraints (IQCs). Finally, the concept is validated in simulation and in a practical experiment using small-scale test vehicles.

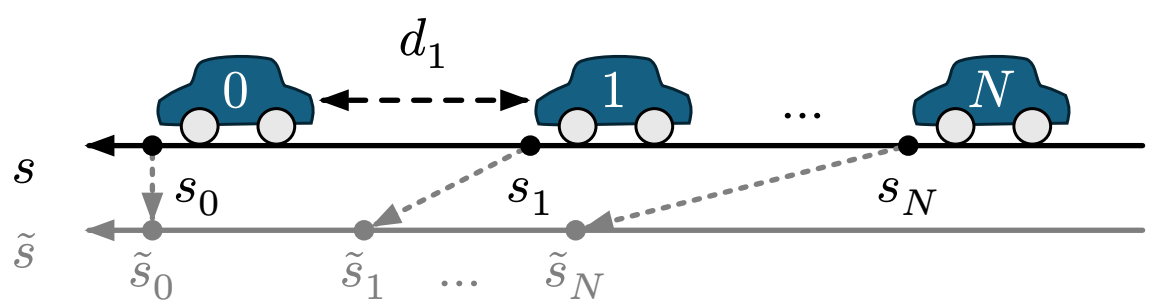

Figure 1. Vehicle platoon of $N$ vehicles with positions shifted by vehicle length and constant distance offset.

### C. *Notation*

Vectors and matrices are written in **bold**, where vectors are written in lower-case and matrices in upper-case. Signals in lower-case letters usually represent time-domain variables, while upper-case letters correspond to Laplace-transformed signals in the complex frequency domain. The expression $\operatorname{diag}(x_1, x_2, ..., x_n)$ denotes a diagonal $n \times n$ matrix with the elements $x_1, x_2, ..., x_n$ on the diagonal. The term *communication delay* refers to the full end-to-end delay which encompasses all components of network-induced delays.

## II. Problem Statement and System Description

In this section, we introduce the system description and control laws for ACC and CACC as presented in, e.g., [2], [3] and [4].

### A. *Vehicle Platoon, System Model and ACC Design*

Consider the vehicle platoon of length $N$ shown in Figure 1. The goal of ACC is the design of a local controller for each vehicle which tracks the velocity of its predecessor while keeping a distance which guarantees collision avoidance. The velocity $v_0(t)$ of the leading vehicle $i = 0$ follows a virtual predecessor which has a given reference velocity $v_{\text{ref}}(t)$. All other following vehicles $i = 1, ..., N$ adapt their velocities $v_i(t)$ to follow their predecessor's velocity $v_{i-1}(t)$ while controlling their distance $d_i(t)$ to keep a given reference distance $d_{i,\text{ref}}(t)$. With the current position $s_i(t)$ of a vehicle with length $l_{V_i}$, the distance to the predecessor and its dynamics are

$$d_i(t) = s_{i-1}(t) - s_i(t) - l_{V_{i-1}} \tag{1}$$

$$\dot{d}_i(t) = v_{i-1}(t) - v_i(t). \tag{2}$$

The desired reference distance

$$d_{i,\text{ref}}(t) = \alpha_i + \beta_i v_i(t), \tag{3}$$

is the sum of a constant offset $\alpha_i$ and a term proportional to the velocity with the time headway $\beta_i$ as proportionality factor. As the distance should never fall below the offset $\alpha_i$, all following investigations are done using the reduced distances

$$\tilde{d}_i(t) = d_i(t) - \alpha_i \tag{4}$$

$$\tilde{d}_{i,\text{ref}}(t) = d_{i,\text{ref}}(t) - \alpha_i = \beta_i v_i(t) \tag{5}$$

which also establishes a linear relation suitable for analysis in frequency domain and leads to the condition

$$\tilde{d}_i(t) \geq 0 \ \forall \ t > 0. \tag{6}$$

Analogously, when reducing the vehicles to point masses and shifting the positions additionally by $\alpha_i$, the shifted positions $\tilde{s}_i$, as depicted in Figure 1, result such that

$$\tilde{d}_i(t) = \tilde{s}_{i-1}(t) - \tilde{s}_i(t). \tag{7}$$

In the beginning of the analysis, the dynamics of a single uncontrolled vehicle $i$ are expressed universally in frequency domain by the transfer function

$$G_i(s) = \frac{V_i(s)}{U_i(s)} \tag{8}$$

with the Laplace transform of the velocity $V_i(s) \bullet\!\!-\!\!\circ\, v_i(t)$ and $U_i(s) \bullet\!\!-\!\!\circ\, u_i(t)$ representing the control input of vehicle $i$ (e.g., the reference acceleration or accelerating force). Based on this, three different local linear ACC controllers are proposed: A purely velocity-based controller which follows the velocity of the preceding vehicle, a distance-based controller which controls the distance to the preceding vehicle, and a velocity- and distance-based controller

$$\begin{aligned} U_i(s) = K_d(s)\big(\beta_i V_i(s) - \tilde{D}_i(s)\big) + \\ K_v(s)(V_{i-1}(s) - V_i(s)) \end{aligned} \tag{9}$$

which combines the two aforementioned ones.

Figure 2 exemplary shows the control structure of the combined distance- and velocity-based controller with pure P-controllers $K_v(s) = k_v$ and $K_d(s) = k_p$. In a practical implementation, $\tilde{d}_i(t)$ is directly measured and fed to the controller $k_p$. For the purpose of system analysis, the control structure includes the dynamics (2) of $\tilde{d}_i(t)$, which depend on the own velocity $v_i(t)$ and the velocity of the preceding vehicle $v_{i-1}(t)$. This allows an analysis as a single-input single-output (SISO) system with $v_{i-1}(t)$ as input instead of a multiple-input system with two inputs $v_{i-1}(t)$ and $\tilde{d}_i(t)$.

Together with definitions (1) – (3), it is possible to determine closed-loop transfer functions of a controlled vehicle for all three controller types as

$$\bar{G}_i(s) = \frac{V_i(s)}{V_{i-1}(s)} \tag{10}$$

$$\bar{F}_i(s) = \frac{\tilde{D}_i(s)}{V_{i-1}(s)} = \frac{1}{s}\Big(1 - \bar{G}_i(s)\Big) \tag{11}$$

with $\tilde{D}_i(s) \bullet\!\!-\!\!\circ\, \tilde{d}_i(t)$, which are used as foundation for further analysis. In particular, a *delay* measure $\Delta_i$ is introduced, which can be used as a measure for the time that $v_i(t)$ takes to adapt to changes in $v_{i-1}(t)$. Figure 3 illustrates an interpretation when vehicle $i$ adapts to a velocity change from $\bar{v}_\text{old}$ to $\bar{v}_\text{new}$ of the preceding vehicle $i-1$.

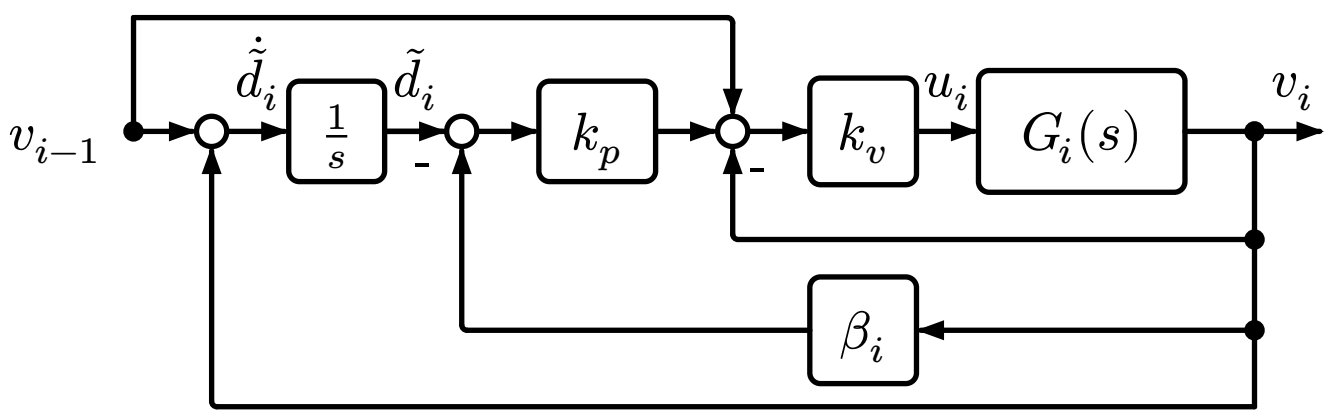

Figure 2. Control structure of combined velocity- and distance controller.

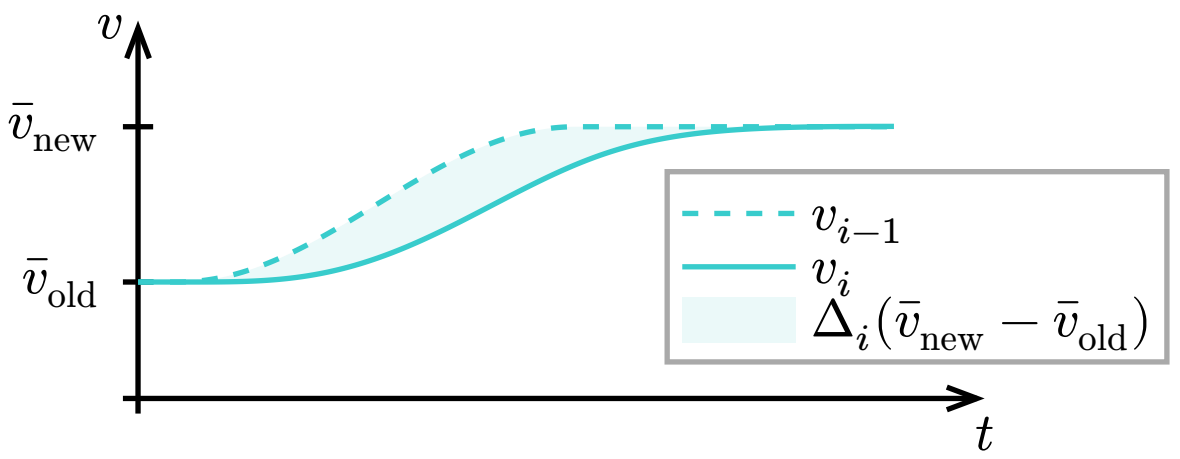

Figure 3. Vehicle $i$ adapting velocity to new velocity of predecessor $i-1$.

Mathematically, this delay is determined by

$$\Delta_i = \bar{F}_i(0) = \int_0^\infty \Big(1 - \bar{h}_i(\tau)\Big)\mathrm{d}\tau \tag{12}$$

where $\bar{h}_i(t) \circ\!\!-\!\!\bullet\, \frac{1}{s}\bar{G}_i(s)$ denotes the step response of the vehicle's velocity transfer function. With that, the following theorem is shown:

**Theorem 2.1** (ACC with asymptotic time headway spacing [1]): The vehicle with the transfer functions (10) follows its predecessor with the requested reference spacing (3) asymptotically, i.e.,

$$\lim_{t\to\infty} \left|\tilde{d}_i(t) - \tilde{d}_{i,\text{ref}}(t)\right| = 0 \tag{13}$$

with initial condition

$$\tilde{d}_i(0) = \beta_i v_i(0) \tag{14}$$

if and only if the delay measure $\Delta_i$ of the controlled vehicle is equal to the time headway $\beta_i$.

Thus, the delay measure $\Delta_i$ determines the minimum distance to the preceding vehicle.

Another important aspect which is shown during the ACC design is that string stability is not a sufficient condition for collision avoidance in all situations [1]. Instead, it is proven that an external positive behavior of the controlled vehicles is required. This condition is fulfilled if positive inputs to the transfer functions (10) lead to positive outputs. Due to its direct relation, it is shown to be sufficient to check this condition for either $\bar{G}_i(s)$ or $\bar{F}_i(s)$.

## B. Extension to CACC

Theorem 2.1 provides a lower bound for the achievable inter-vehicle distances in ACC. In order to realize smaller time headway factors $\beta_i < \Delta_i$ than the vehicle delay allows, information of vehicles further in front is required. The solution presented in [3] proposes to replace the velocity of the preceding vehicle as reference velocity by a weighted sum of vehicle velocities further in front of the ego-vehicle

$$v_\text{si}(t) = \sum_{j=0}^{i} \hat{a}_{i,j} v_j(t) \quad \text{with} \quad \sum_{j=0}^{i} \hat{a}_{i,j} = 1 \tag{15}$$

where the velocities $v_j(t)$ for $j < i-1$ are not locally measurable and must be sent via V2V communication. A systematic approach to determine the weighting factors $\hat{a}_{i,j}$ is presented and it is shown to be sufficient to establish communication links to at most two vehicles in front, whose indices are referred to as $l_i$ and its predecessor $l_i - 1$. The procedure of determining $l_i$ and the weighting factors $\hat{a}_{i,l_i}$ and $\hat{a}_{i,l_i-1}$ is summarized in Theorem 2.2.

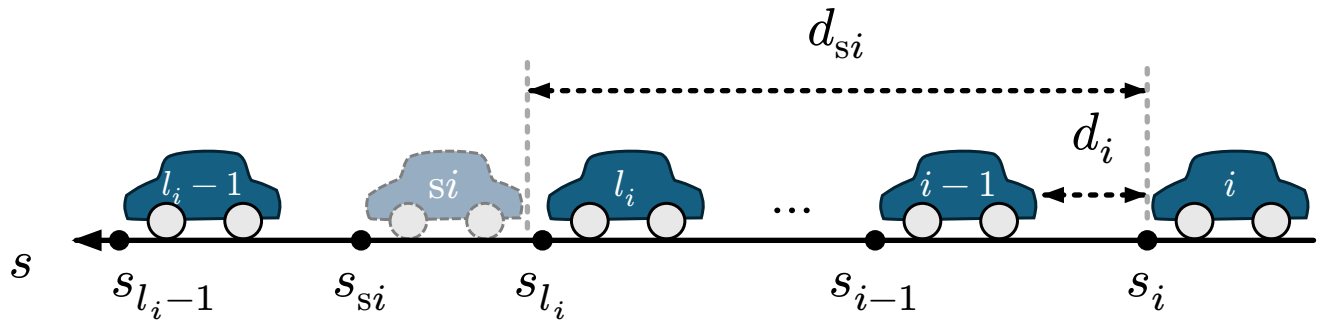

Figure 4. Control to virtual predecessor in CACC.

**Theorem 2.2** (Design of Communication Structure for CACC [2], [3]): Consider the vehicle platoon in Figure 1. An asymptotic time headway spacing $\beta_i < \Delta_i$ for vehicle $i$ can be achieved by choosing the index $l_i$ according to the condition

$$0 \leq \sum_{j=l_i}^{i} \beta_j - \Delta_i \leq \beta_{l_i}, i = 2, 3, ..., N \tag{16}$$

and the corresponding coefficients

$$\begin{aligned}
\hat{a}_{1,0} &= 1 \\
\hat{a}_{i,l_i} &= \frac{\sum_{j=l_i}^{i} \beta_j - \Delta_i}{\beta_{l_i}}, i = 2, 3, ..., N \\
\hat{a}_{i,l_i-1} &= 1 - \hat{a}_{i,l_i} \qquad , i = 2, 3, ..., N \\
\hat{a}_{i,j} &= 0 \qquad\qquad , j \notin \{l_i, l_i - 1\}
\end{aligned} \tag{17}$$

to determine the networked reference velocity $v_\text{si}(t)$ from (15).

Figure 4 shows an interpretation of the networked reference velocity $v_\text{si}(t)$: The communication structure is designed such that vehicle $i$ follows a virtual preceding vehicle (denoted by the subscript s$i$) with time headway factor $\Delta_i$, analogously to choosing $\beta_i = \Delta_i$ in ACC, while asymptotically keeping the smaller reference distance with time headway factor $\beta_i < \Delta_i$ to the direct predecessor $i-1$. For a formal proof of collision avoidance, external positivity of a further transfer function

$$G_\text{di}(s) = \frac{\tilde{D}_i(s)}{V_0(s)} \tag{18}$$

describing the relation between the reduced distances and the leader's velocity must be shown. The derivation and construction of this transfer function can be found in [3] and [4].

## C. Control of the Distance in CACC

An important aspect to highlight, additionally to the existing investigations, is that the replacement of $v_{i-1}(t)$ by $v_{\mathrm{s}i}(t)$ in a controller leads to changes when using a controller with distance part. Consider, e.g., the controller in Figure 2 that controls the velocity and distance. Replacing the input $v_{i-1}(t)$ by $v_{\mathrm{s}i}(t)$ and keeping the remaining control diagram unchanged does now describe the reduced distance to the **virtual** predecessor, which we call $\tilde{d}_{\mathrm{s}i}(t)$, rather than the reduced distance to the direct predecessor $\tilde{d}_i(t)$. Three possibilities to handle this situation exist:

1) Use a purely velocity-based controller, which does not have a distance-feedback at all.
2) Keep the input $v_{i-1}(t)$ for the distance controller part and only replace $v_{\mathrm{s}i}(t)$ for the velocity-based part. This however leads to a multiple-input system where the developed theory cannot be applied without further investigations.
3) Control the reduced distance to the virtual predecessor $\tilde{d}_{\mathrm{s}i}(t)$ and replace the feedback factor $\beta_i$ in Figure 2 by $\Delta_i$. By design, this keeps the theory valid and indirectly controls the distance to the direct predecessor asymptotically to its reference.[1]

As the interpretation of the virtual distance is only given once in [4] without highlighting the required changes that would be necessary when controlling it, we assume that the first option without any distance control is the one used in the underlying design procedure. Since a distance feedback control in some form is necessary in real practical applications to allow adjustments of the desired reference and compensate for initial offsets, it is not a practical solution. To keep the preceding theory valid, we continue with the third option. The drawback of this option becomes clear from Figure 4: for vehicle $i$, two communication links to the vehicles $l_i$ and $l_i - 1$ may not suffice anymore. Instead, vehicle $i$ must receive the measured distances $\tilde{d}_j(t)$ from all vehicles $j = l_i, ..., i-1$ via V2V communication to calculate the distance to the virtual predecessor $\tilde{d}_{\mathrm{s}i}(t)$. In most practical setups, this is not a huge constraint as the vehicles to be communicated with are usually not too far ahead. As an example, it is shown in [2] that for a homogeneous platoon with identical vehicles and identical time headway, only two communication links to the direct two predecessors are required. In this case, the number of communication links does not even increase when controlling $\tilde{d}_{\mathrm{s}i}(t)$.

Using the shifted positions $\tilde{s}$ from Figure 1, it is

$$\begin{aligned}\dot{\tilde{d}}_{\mathrm{s}i}(t) &= v_{\mathrm{s}i}(t) - v_i(t) \\ &= \hat{a}_{i,l_i}\dot{\tilde{s}}_{l_i}(t) + \hat{a}_{i,l_i-1}\dot{\tilde{s}}_{l_i-1}(t) - \dot{\tilde{s}}_i(t).\end{aligned} \tag{19}$$

Integrating (19) and including relations between the shifted positions yields

$$\tilde{d}_{\mathrm{s}i}(t) = \sum_{j=l_i+1}^{i} \tilde{d}_j(t) + \hat{a}_{i,l_i-1}\tilde{d}_{l_i}(t) \tag{20}$$

which underlines the previous statement that communication from all vehicles up to $l_i - 1$ is required.

# III. Time-Varying Delays and Parameters

All results from the previous section apply to linear time-invariant (LTI) systems. As mentioned in [3] and [4], a communication delay $\tau_i$ between the vehicles can be included in the delay measure $\Delta_i$ of a controlled vehicle and the design procedure remains valid in the time-invariant case. However, in practice, the communication delay is not constant over time. Therefore, we propose to dynamically adapt the overall delay

$$\hat{\Delta}_i(t) = \Delta_i + \tau_i(t) \tag{21}$$

online when the QoS of the communication network changes. This sum appears in the design procedure of Theorem 2.2 and the feedback control of the virtual distance (20) which should now be controlled to

$$\tilde{d}_{\mathrm{s}i,\mathrm{ref}}(t) = \hat{\Delta}_i(t)v_i(t). \tag{22}$$

Thus, it also appears in the closed-loop system of the controlled vehicle. As stability of each controlled vehicle is a necessary condition, we prove it for an exemplary LPV vehicle model with variable time delays as a first step.

As further consequence, it may become impossible to keep the desired time headway $\beta_i$ under severely deteriorating QoS. We therefore analyze the influence of an online adaptation of $\beta_i(t)$. While it is possible to keep the factor constant in a limited range by adjusting the communication topology, it must be enlarged when the overall delay $\hat{\Delta}_i(t)$ increases above the limit where the condition in Theorem 2.2 does not have a solution anymore, or when the connection is completely lost. Hence, we replace the constant factor $\beta_i$ by a time-varying parameter $\beta_i(t)$ and highlight the necessary changes in the deployment algorithm.

While the remaining analysis will be valid for arbitrarily fast variations of $\hat{\Delta}_i(t)$ and $\tau_i(t)$, a practical implementation requires an advanced network monitoring system that adapts the time headway factor in (22) only for long-term changes of $\tau_i(t)$, not for rapid jitter. First investigations on a similar use case with another controller are done in [10], but such a policy depends on the controller and further assumptions on the network. Developing a policy for our controller is beyond the scope of this work, but is subject to further research.

[1] Even though the distance to the direct predecessor $\tilde{d}_i(t)$ is part of $\tilde{d}_{\mathrm{s}i}(t)$, it is not controlled directly. Large disturbances in other parts of $\tilde{d}_{\mathrm{s}i}(t)$ can therefore lead to unexpected behavior of $\tilde{d}_i(t)$. We assume that all vehicles reliably control $\tilde{d}_{\mathrm{s}i}(t)$ in the desired manner, but this fact should be kept in mind and subject of future research.

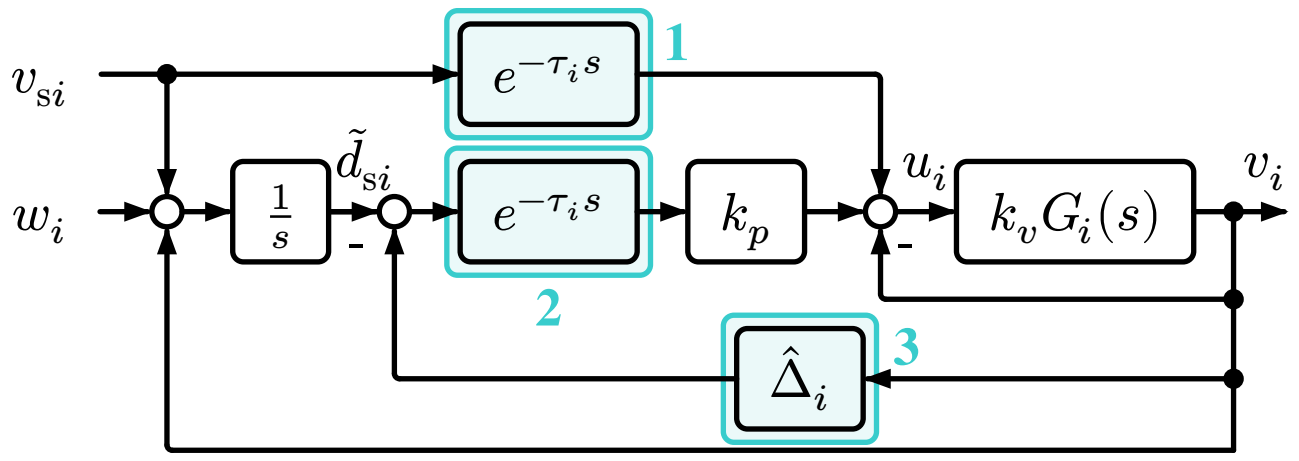

Figure 5. Extended control diagram of combined velocity- and distance controller including communication delays and adapted time headway.

### A. System Description in Time Domain

In the following, we consider the commonly used system model for the uncontrolled vehicle

$$G_i(s) = \frac{1}{s}\left(\frac{1}{Ts+1}\right) \tag{23}$$

which represents a simple integrator dynamics between the vehicle velocity $v_i(t)$ and its acceleration $a_i(t)$. The first-order lag element with time constant $T$ models the input lag of the drivetrain between $a_i(t)$ and the reference acceleration $u_i(t)$. As explained in the previous section, we focus on controlling the distance to the virtual predecessor $\tilde{d}_{si}(t)$ to its reference (22). The description in time-domain of the system is then

$$\underbrace{\begin{pmatrix} \dot{v}_i(t) \\ \dot{a}_i(t) \\ \dot{\tilde{d}}_{si}(t) \end{pmatrix}}_{\dot{\boldsymbol{x}}_i(t)} = \underbrace{\begin{pmatrix} 0 & 1 & 0 \\ 0 & -\frac{1}{T} & 0 \\ -1 & 0 & 0 \end{pmatrix}}_{\boldsymbol{A}_i} \underbrace{\begin{pmatrix} v_i(t) \\ a_i(t) \\ \tilde{d}_{si}(t) \end{pmatrix}}_{\boldsymbol{x}_i(t)} + \underbrace{\begin{pmatrix} 0 \\ \frac{1}{T} \\ 0 \end{pmatrix}}_{\boldsymbol{B}_i} u_i(t) + \underbrace{\begin{pmatrix} 0 \\ 0 \\ 1 \end{pmatrix}}_{\boldsymbol{e}_i} v_{si}(t) \tag{24}$$

for the uncontrolled vehicle. Using the proposed velocity- and distance-based controller (9) with replacements $V_{i-1}(s) \to V_{si}(s)$, $\tilde{D}_i(s) \to \tilde{D}_{si}(s)$ and $\beta_i \to \Delta_i$, and with proportional controllers $K_v(s) = k_v$ and $K_d(s) = k_p$, the time-domain formulation for controlling the distance to the virtual predecessor

$$u_i(t) = k_v(v_{si}(t) - v_i(t) + k_p(\Delta_i v_i(t) - \tilde{d}_{si}(t)) \tag{25}$$

results with the state-space model for the controlled vehicle

$$\dot{\boldsymbol{x}}_i(t) = \begin{pmatrix} 0 & 1 & 0 \\ \frac{-k_v + k_v k_p \Delta_i}{T} & -\frac{1}{T} & -\frac{k_v k_p}{T} \\ -1 & 0 & 0 \end{pmatrix} \boldsymbol{x}_i(t) + \begin{pmatrix} 0 & \frac{k_v}{T} & 1 \end{pmatrix}^{\mathsf{T}} v_{si}(t). \tag{26}$$

To further include communication delays $\tau_i(t)$, it is necessary to (i) substitute $\Delta_i$ by $\hat{\Delta}_i(t)$ from (21), (ii) replace $v_{si}(t)$ by $v_{si}(t - \tau_i(t))$, and (iii) include a time delay for the measurements of $\tilde{d}_{si}(t)$.[2]

These steps turn the LTI system into an LPV system with variable time delays. Figure 5 shows the extension of Figure 2 to the time-varying system. The differences between those control structures are derived and explained in Section III.B and Section III.C.

### B. Communication Delays of Velocities and Distances

The blue boxes in Figure 5 with labels 1 and 2 represent the communication delays of $v_{si}(t)$ and $\tilde{d}_{si}(t)$, respectively. Note that the signal $\tilde{d}_{si}(t)$ represents the real physical value. As a consequence, the communication delay block of $v_{si}(t)$ is inserted *after* the branch of $v_{si}(t)$ to the integrator of $\tilde{d}_{si}(t)$ to keep the physical relation between $v_{si}(t)$ and $\tilde{d}_{si}(t)$ valid. For placing the communication delay of $\tilde{d}_{si}(t)$, two options exist: (i) directly in the signal path of $\tilde{d}_{si}(t)$, where it physically results from, or (ii) in the signal path of the distance control error $e_d(t)$. When the communication structure is updated immediately according to Theorem 2.2 when $\tau_i(t)$ changes, the second option does not lead to any changes in $e_d(t)$ under ideal circumstances. In contrast, this would not be the case for the first option as a result of different time delays between the measured value $\tilde{d}_{si}(t)$ and its reference. Consequently, we choose the second option, which requires manually implementing an additional time delay of the feedback term (22).

Lastly, the time headway factor for the virtual distance in (22) is to be adapted to $\hat{\Delta}_i(t)$. It must be shown that the overall delay (12) is still defined by this factor in steady state for a constant $\tau_i$. For this case, the transfer functions $\bar{G}_i(s)$ and $\bar{F}_i(s)$ from (10) and (11) for model (23) can be derived as

$$\bar{G}_i(s) = \frac{k_v(s - k_d)e^{-\tau_i s}}{Ts^3 + s^2 + k_v s - k_v k_d e^{-\tau_i s}(\hat{\Delta}_i s + 1)} \tag{27}$$

$$\bar{F}_i(s) = \frac{Ts^2 + s - k_v k_d \hat{\Delta}_i e^{-\tau_i s}}{Ts^3 + s^2 + k_v s - k_v k_d e^{-\tau_i s}(\hat{\Delta}_i s + 1)}, \tag{28}$$

and using definition (12), the overall vehicle delay is defined by the feedback term of the distance control loop

$$\bar{F}_i(0) = \hat{\Delta}_i = \Delta_i + \tau_i \tag{29}$$

which proves that the distance control to the virtual predecessor can be applied despite the communication delays.

### C. Recalculation of the Communication Coefficients

The communication structure and coefficients depend on the overall delay $\hat{\Delta}_i(t)$ of each vehicle (see (17) in Theorem 2.2). We assume a time-invariant communication structure, i.e., $l_i$ is kept constant.[3] Whenever $\hat{\Delta}_i(t)$ changes, condition (16) must at first be evaluated. If $\hat{\Delta}_i(t)$ exceeds a limit such that it is violated, the time headway $\beta_i(t)$ must be increased dynamically until (16) is fulfilled. Vice versa, $\beta_i(t)$ may be decreased when $\hat{\Delta}_i(t)$ reduces. After satisfying (16), the coefficients $\hat{a}_{i,l_i}$ and $\hat{a}_{i,l_i-1}$ are recalculated. The resulting change of the reference velocity does not influence the control system structure. However, the virtual distance (20) depends on $\hat{a}_{i,l_i-1}$, which does influence the system model. Differentiating (20) again, this time with time-varying $\hat{a}_{i,l_i-1}(t)$, yields

$$\dot{\tilde{d}}_{si}(t) = v_{si}(t) - v_i(t) + \dot{\hat{a}}_{i,l_i-1}(t) \cdot \tilde{d}_{l_i}(t) \tag{30}$$

where the last term now appears additionally compared to (19) in the nominal case. This new term can be considered as external disturbance input

$$w_i(t) := \dot{\hat{a}}_{i,l_i-1}(t) \cdot \tilde{d}_{l_i}(t) \tag{31}$$

which is added in Figure 5. Due to the dependency on the time headway factors of the vehicles ahead, all vehicles behind vehicle $i$, i.e., all vehicles $j = i+1, ..., N$, must be notified when a change in $\beta_i(t)$ is executed. These vehicles must then check whether they have to redefine their communication.

[2]Remember that the dynamics of the state $\tilde{d}_{si}(t)$ are included in the systems (24) and (26) for analysis. In practice, $\tilde{d}_{si}(t)$ is also measured and communicated with a time delay.

[3]A time-varying $l_i(t)$ leads to a time-varying lower bound in the sum (20) whose influence requires further research.

This deployment and adaptation algorithm of one vehicle is shown in the flow chart of Figure 6.

## IV. Robustness Analysis with Integral Quadratic Constraints

The system described above is an LPV system with time delays. As the well-known robustness analysis tool $\mu$-analysis only makes statements about bounded, but time-invariant parameters, an alternative must be used. In the following, we use the IQC framework. J. Veenman *et al.* [11] developed a MATLAB-toolbox *IQClab*, which is an extension of MATLAB's own Robust Control Toolbox, providing an easy-to-use interface for several IQC-based robustness analysis methods. We only use the tools provided by IQC. Since an introduction to this is beyond the scope of this work, we refer to existing literature, e.g., [11], [12] and [13]. The remainder of this section explains the uncertainty descriptions for the IQC analysis.

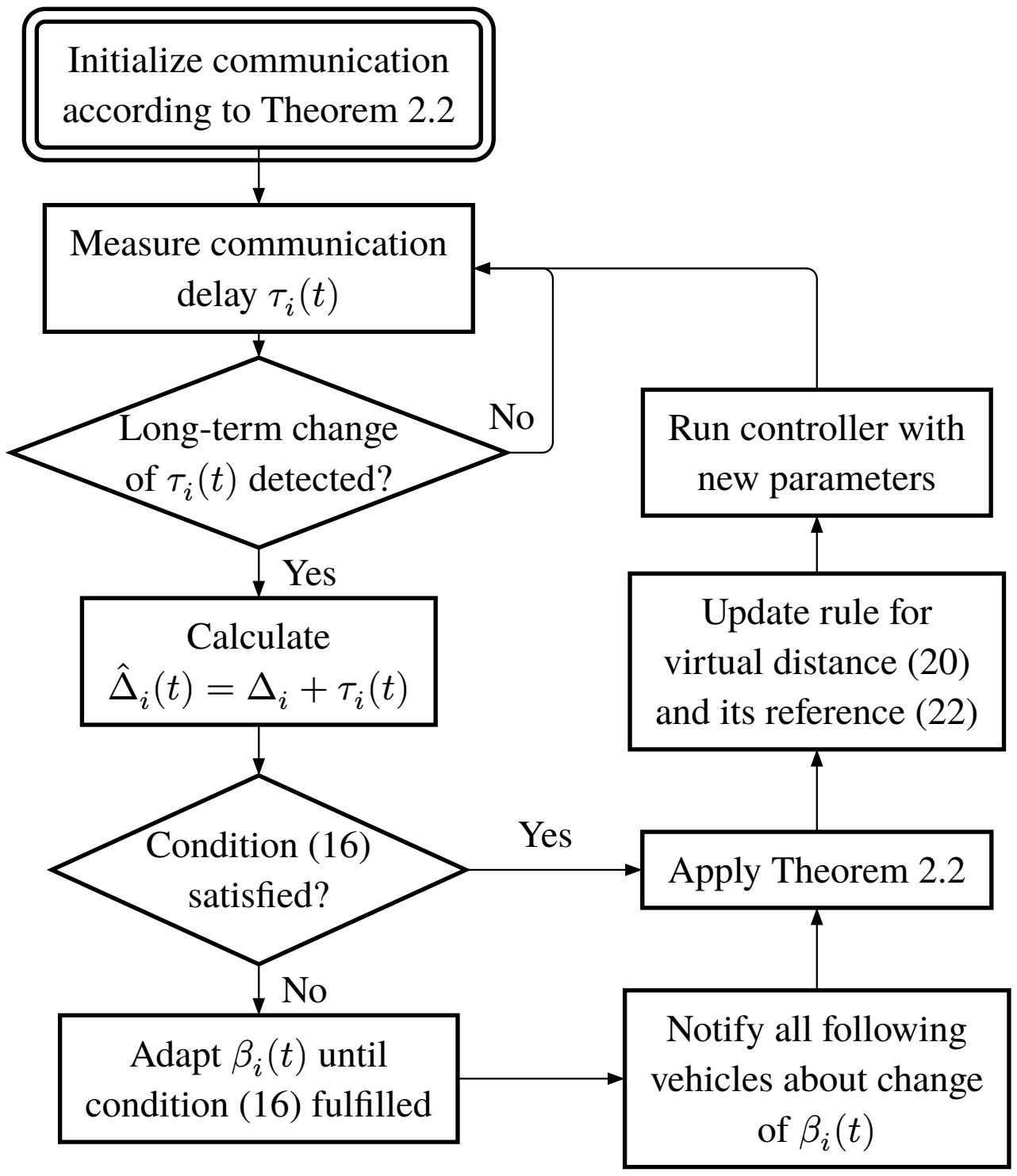

Figure 6. Flow chart of the online-algorithm to monitor the current communication delays and adapt the controller structure accordingly.

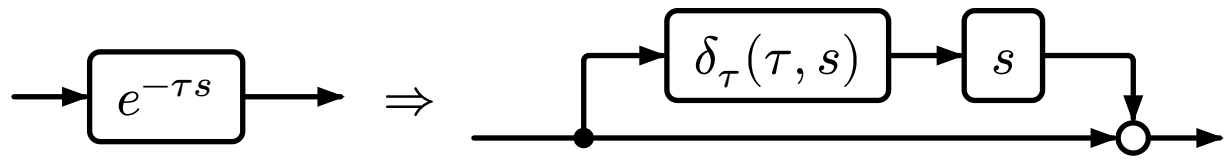

Figure 7. Reformulation of the delay in control diagram for IQC analysis.

### A. Uncertainty-Descriptions of the Virtual Time Headway and Communication Delays

As usual in robust control theory, the time-varying parameter $\hat{\Delta}_i(t)$ in (21) is modeled as bounded uncertainty of the form

$$\hat{\Delta}_i(t) = \bar{\Delta}_i\big(1 + r_{\Delta_i}\delta_{\Delta_i}(t)\big) \tag{32}$$

with the nominal value $\bar{\Delta}_i$, the norm-bounded time-varying uncertainty $\delta_{\Delta_i}(t) \in [-1, 1]$ and the scaling factor

$$r_{\Delta_i} = \frac{\hat{\Delta}_{i_{\max}} - \hat{\Delta}_{i_{\min}}}{\hat{\Delta}_{i_{\max}} + \hat{\Delta}_{i_{\min}}} \tag{33}$$

depending on the minimum and maximum bounds on $\hat{\Delta}_i(t)$. In IQC analysis, the uncertainty $\delta_{\Delta_i}(t)$ is modeled as *arbitrarily fast time-varying parametric uncertainty*.

Concerning uncertainties in time delays, two different classes are presented in [12] and implemented in *IQCLab*. Only one of them can be used for arbitrarily fast time-varying time delays. In this class, the uncertainty is defined as

$$\delta_\tau(\tau, s) = s^{-1}(e^{-\tau s} - 1) \tag{34}$$

which requires restructuring the time-delay blocks $e^{-\tau s}$ with labels 1 and 2 in Figure 5 to keep the same I/O-behavior as shown in Figure 7. The differentiation operator $s$ leads to a non-causal system when setting up the system model. To overcome this issue for system analysis, the state-space representation of $G_i(s)$ in (23) can be modified such that the first state is just an integration of the control input. That way, the output of the uncertainty can be shifted so that it is added after this integrating state and the differentiation operator $s$ is canceled with the integrator. Figure 8 illustrates the plant state transformation and the shift.

### B. Choice of the Control Parameters

In the delay-constant case, it is shown that

$$k_v = 1/\Delta_i \tag{35}$$

results in the desired delay for a pure proportional velocity-controller [1]. Further simulations have shown that keeping this value constant and not adapting it to $1/\hat{\Delta}_i(t)$ is the superior choice for the time-varying case. The control parameter $k_p$ should be selected such that the external behavior of (27) is kept for any stationary $\tau_i \in \big[0, \tau_{i,\max}\big]$. Here, a trade-off is necessary: A higher value of $\tau_i$ requires a lower $k_p$. However, the smaller $k_p$, the slower the distance control loop and the regulation process when changing the (virtual) time headway.

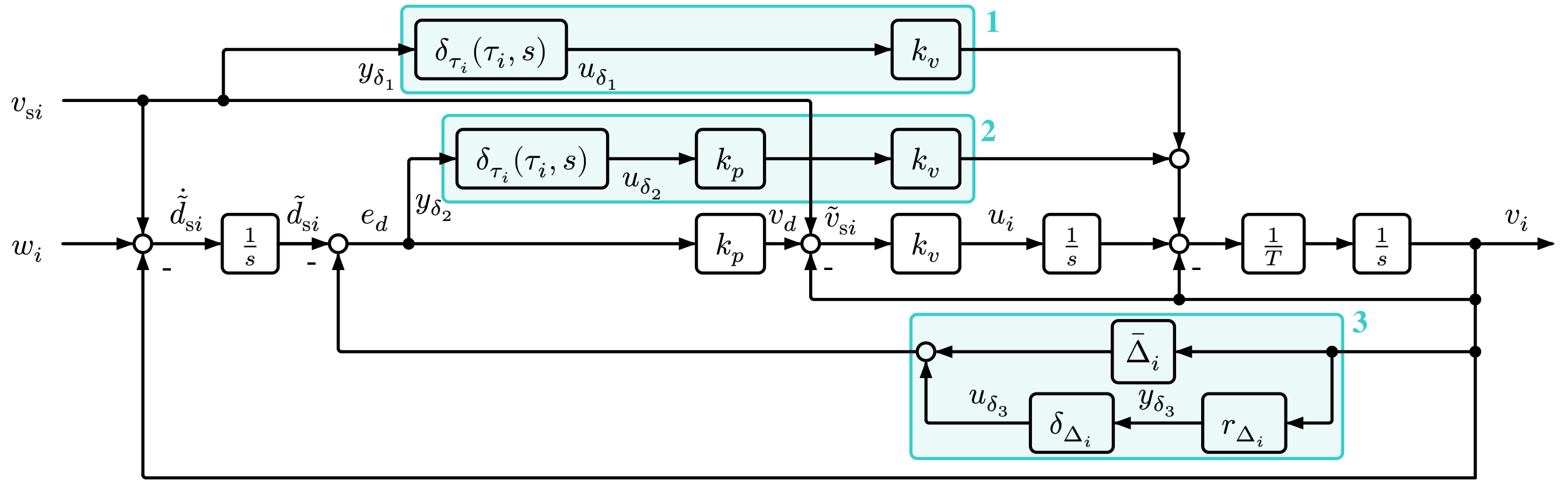

Figure 8. Control diagram for IQC robustness analysis. Boxes 1 and 2 illustrate the inclusion of the uncertain communication delay operators after restructuring for a causal system model. Box 3 illustrates the uncertainty model of the virtual time headway feedback factor.

### C. State-Space Model of the Uncertain System

After having introduced all changes arising in the time-varying system, the state-space model of Figure 8 can be set up. Rearranging the blocks yields a block-diagonal uncertainty matrix

$$\boldsymbol{\delta}(t) = \mathrm{diag}\big(\delta_{\tau_i}(t), \delta_{\tau_i}(t), \delta_{\Delta_i}(t)\big) \tag{36}$$

which allows defining

$$\boldsymbol{u}_{\boldsymbol{\delta}}(t) = \begin{pmatrix} u_{\delta_1}(t) & u_{\delta_2}(t) & u_{\delta_3}(t) \end{pmatrix}^\mathsf{T} \tag{37}$$

$$\boldsymbol{y}_{\boldsymbol{\delta}}(t) = \begin{pmatrix} y_{\delta_1}(t) & y_{\delta_2}(t) & y_{\delta_3}(t) \end{pmatrix}^\mathsf{T} \tag{38}$$

such that

$$\boldsymbol{u}_{\boldsymbol{\delta}}(t) = \boldsymbol{\delta}(t)\boldsymbol{y}_{\boldsymbol{\delta}}(t). \tag{39}$$

Finally, with the definition of $\boldsymbol{x}_i(t)$ in (24) and the extended control input vector for the robustness analysis

$$\boldsymbol{u}_\mathrm{r}(t) = \begin{pmatrix} v_{\mathrm{s}i}(t) & w_i(t) & \boldsymbol{u}_{\boldsymbol{\delta}}(t) \end{pmatrix}^\mathsf{T} \tag{40}$$

the state-space model

$$\begin{aligned} \dot{\boldsymbol{x}}_i(t) &= \boldsymbol{A}_\mathrm{r}\boldsymbol{x}_i(t) + \boldsymbol{B}_\mathrm{r}\boldsymbol{u}_\mathrm{r}(t) \\ \boldsymbol{y}_{\boldsymbol{\delta}}(t) &= \boldsymbol{C}_\mathrm{r}\boldsymbol{x}_i(t) + \boldsymbol{D}_\mathrm{r}\boldsymbol{u}_\mathrm{r}(t) \end{aligned} \tag{41}$$

with the matrices

$$\boldsymbol{A}_\mathrm{r} = \begin{pmatrix} -\frac{1}{T} & \frac{1}{T} & 0 \\ k_v k_p \bar{\Delta}_i - k_v & 0 & -k_p k_v \\ -1 & 0 & 0 \end{pmatrix} \quad \boldsymbol{C}_\mathrm{r} = \begin{pmatrix} 0 & 0 & 0 \\ \bar{\Delta}_i & 0 & -1 \\ r_{\Delta_i} & 0 & 0 \end{pmatrix}$$

$$\boldsymbol{B}_\mathrm{r} = \begin{pmatrix} 0 & 0 & \frac{k_v}{T} & \frac{k_v k_p}{T} & 0 \\ k_v & 0 & 0 & 0 & k_v k_p \\ 1 & 1 & 0 & 0 & 0 \end{pmatrix} \quad \boldsymbol{D}_\mathrm{r} = \begin{pmatrix} 1 & 0 & 0 & 0 & 0 \\ 0 & 0 & 0 & 0 & 1 \\ 0 & 0 & 0 & 0 & 0 \end{pmatrix}$$

results. Applying the methods in the *IQClab* toolbox to this system is straightforward. After identifying $T$ and defining $k_p$, $k_v$ and the uncertainty bounds, the state-space model (41) is implemented in MATLAB. The uncertainty matrix $\boldsymbol{\delta}(t)$ can be defined as uncertainty object within its bounds and assigned to the input channels of $\boldsymbol{u}_{\boldsymbol{\delta}}(t)$. Executing the IQC analysis solves a linear matrix inequality (LMI) optimization problem. Feasibility of the LMI problem establishes robust stability.

## V. Simulation and Practical Results

This section provides simulation results and practical evaluations on small-scale test vehicles to verify that controlling the distance to the virtual predecessor works as expected and to analyze the influence in dynamic QoS, reflected by changes in the communication delays and in turn the time headway. As a first step, the time constant $T$ of the test vehicles was experimentally identified as $T = 0.3\,\mathrm{s}$ with which a vehicle delay $\Delta_i = 1.2\,\mathrm{s}$ can be achieved in the case of no communication delays. The maximal communication delay is assumed to be $\tau_\mathrm{max} = 0.6\,\mathrm{s}$. With that, the controller parameters $k_v$ and $k_p$ were designed to fulfill the condition on external positivity for all time-stationary $\tau(t) = \tau \leq \tau_\mathrm{max}$ by evaluating the step responses of (27) numerically for a sufficiently dense grid of $\tau$. Applying these parameters to the IQC-based robustness analysis from Section IV proves robust stability.

### A. Simulation Results

The simulations are executed using MATLAB/Simulink with a platoon of four vehicles with the same parameters as the test vehicles. The scenario includes a communication delay change without influence on the platoon behavior as well as a case requiring a time headway increase of one vehicle. The leading vehicle follows a step-shaped reference trajectory $v_\mathrm{ref}$. The first following vehicle ($i = 1$) receives the velocity of the leader via wireless communication with a constant delay $\tau_1 = 0.1\,\mathrm{s}$ and thus follows with a time headway $\beta_1 = \Delta_1 + \tau_1 = 1.3\,\mathrm{s}$. All following vehicles operate in CACC with $l_i = i - 1$ and a desired $\beta_{i,\mathrm{des}} = 0.8\,\mathrm{s}$. The time-varying communication delays for these vehicles in CACC mode are shown in Figure 9, together with the simulation results of the velocities, reduced distances and adaptive time headways.

Notably, the first increase of the communication delays $\tau_2(t)$ and $\tau_3(t)$ at $t = 10\,\mathrm{s}$ does not have any qualitative effect on the platoon behavior because this change can be caught by a recalculation of the communication parameters $\hat{a}_{i,j}(t)$. In contrast, this is not possible after the second increase at $t = 20\,\mathrm{s}$.

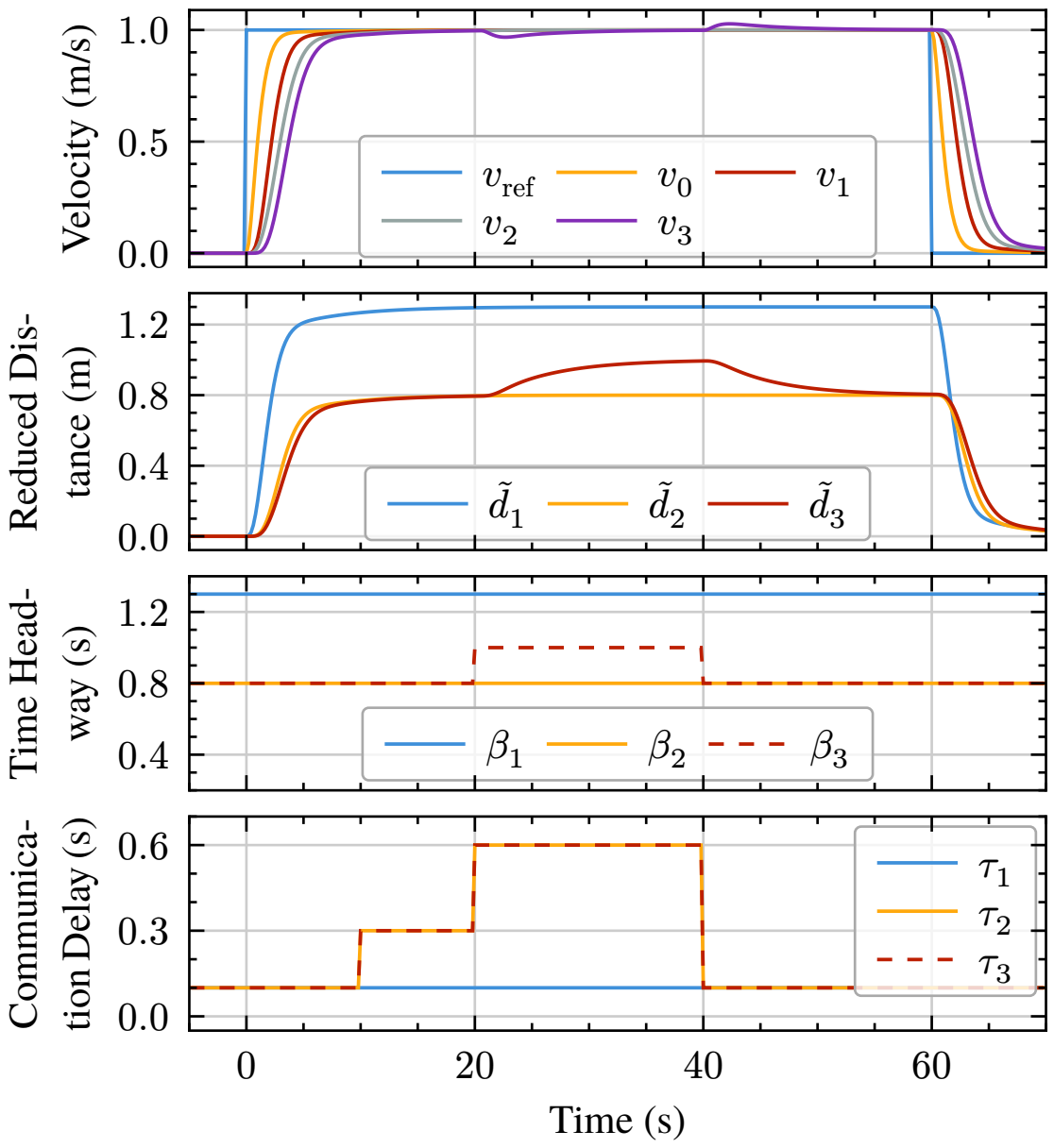

Figure 9. Vehicle velocities, reduced distances, time headway factors and communication delays in simulation.

As condition (16) is violated for vehicle 3, it dynamically increases its time headway factor $\beta_3(t)$ such that the condition is fulfilled again, which can be seen by the increasing reduced distance $\tilde{d}_3(t)$. After $t = 40$ s, the desired time headway can be established again and the distance decreases.

### B. Experimental Setup and Results

The experimental setup consists of four test vehicles in a scale of 1:10 as shown in Figure 10. The controller parameters, desired time headways and communication delays (which are introduced artificially for validation) match the ones in the simulation. Figure 11 depicts the measurements. Analogously to the simulation, the dynamic distance adaptation at $t = 20$ s can be observed. While the small-scaled testbed omits effects of real vehicles (e.g., missing air- and rolling resistances or different powertrain dynamics), the similarity to the simulation confirms the qualitative feasibility of the distance control to the virtual predecessor and the adaptive time headway.

## VI. Conclusion

Extending an existing CACC design, we presented a graceful degradation strategy by dynamically adjusting the inter-vehicle distances based on the current network QoS. We highlighted the necessary steps for this adjustment in the controller design, introduced a distance control to a virtual predecessor, and proved IQC-based stability of the resulting LPV system with variable communication delays. By design, this ensures collision avoidance in steady state (i.e., when the distances errors are regulated to zero for a constant communication delay). Future work may address collision avoidance analysis during time headway transients, an extension to time-varying communication structures and network monitoring policies that distinguish short- and long-term QoS changes.

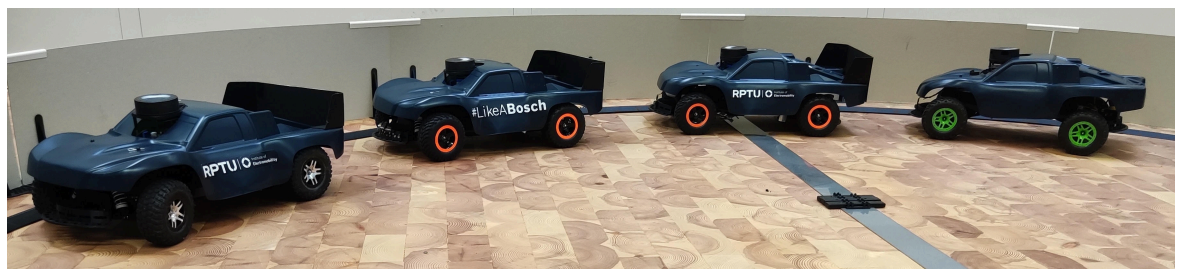

Figure 10. Experimental test vehicle setup.

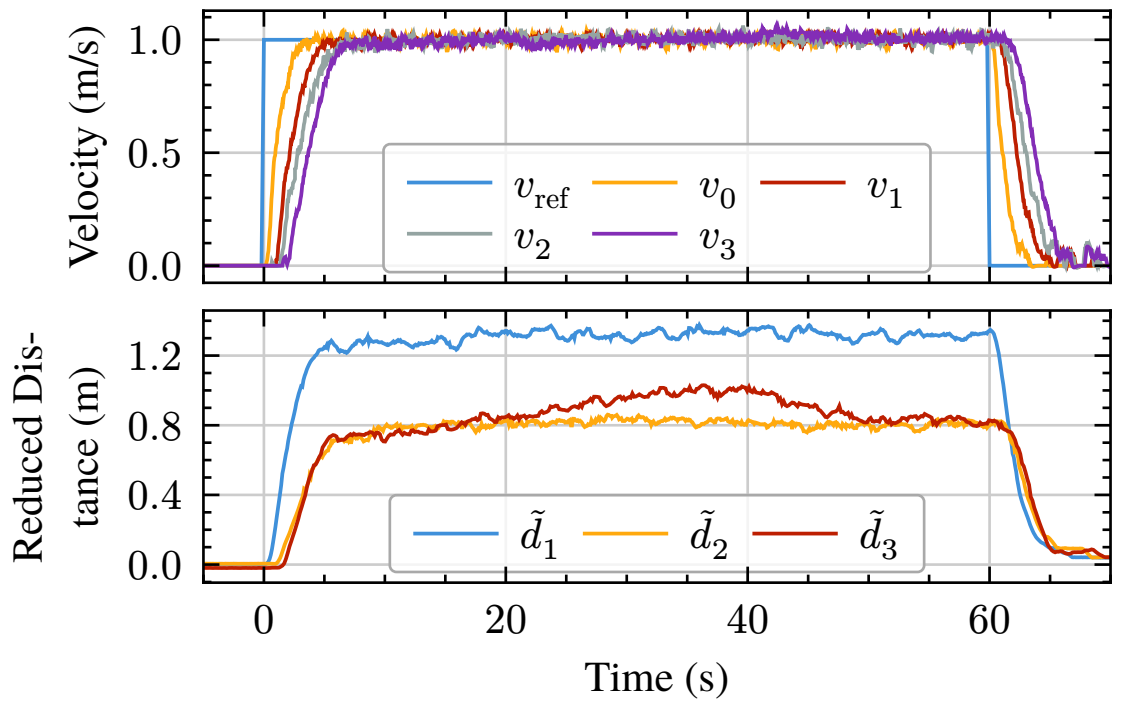

Figure 11. Vehicle velocities and reduced distances in experiment.